\documentclass[prl,preprint,showpacs,floatfix]{revtex4}

\usepackage{amsmath}
\usepackage{epsfig,amssymb,txfonts} % txfonts uses times throughout
\usepackage{amssymb}
\usepackage{amsfonts}

\newcommand*{\be}[0]{\begin{equation}}
\newcommand*{\ee}[0]{\end{equation}}
\newcommand*{\beu}[0]{\begin{equation*}}
\newcommand*{\eeu}[0]{\end{equation*}}
\newcommand*{\ba}[0]{\begin{array}}
\newcommand*{\ea}[0]{\end{array}}
\newcommand*{\bfig}[0]{\begin{figure}[t]} %[!ht]
\newcommand*{\efig}[0]{\end{figure}}
\newcommand*{\bfigwide}[0]{\begin{figure*}}
\newcommand*{\efigwide}[0]{\end{figure*}}

\newlength{\wholefigwidth}
\newlength{\smallfigwidth}
\newlength{\halfsmallfigwidth}
\newcommand{\Fig}[1]{Fig.~\ref{fig:#1}}

\newcommand{\eps}[0]{\varepsilon}

\begin{document}

%%%%%%%%%%%%%%%%%%%%%%%%%%%%%%%%%%%%%%%%%%%%%%%%%%%
\title{Island Shape Controls Magic-Size Effect for Heteroepitaxial Diffusion}

\author{Henry H. Wu}
\author{A. W. Signor}
\author{Dallas R. Trinkle}
\email{dtrinkle@illinois.edu}
\affiliation{Department of Materials Science and Engineering, University of
Illinois, Urbana-Champaign}

\date{\today}
\begin{abstract}
Lattice mismatch of Cu on Ag(111) produces fast diffusion for special
``magic sizes'' of islands.  A size- and shape-dependent reptation
mechanism is responsible for low diffusion barriers.  Initiating the
reptation mechanism requires a suitable island shape, a property not
considered in previous studies of 1D island chains and 2D closed-shell
islands.  Shape determines the dominant diffusion mechanism and leads to
multiple clearly identifiable magic-size trends for diffusion depending on
the number of atoms whose bonds are shortened during diffusion.
\end{abstract}
% Diffusion, Metal Alloy, Surface Strain, Molecular dynamics

\maketitle

%%%%%%%%%%%%%%%%%%%%%%%%%%%%%%%%%%%%%%%%%%%%%%%%%%%%%%%%%%%%%%%%%%%%%%%%

Control of thin-film morphology relies on understanding multiple ongoing
processes during deposition and growth.  In particular, diffusion of small
atom clusters on surfaces play a critical role in thin film growth,
especially in early stages.  The diffusion kinetics of small islands in
heteroepitaxial systems is less well understood than that of homoepitaxial
diffusion, for which much experimental%
\cite{Wen:1994uq,Kellogg:1991yq,Bartelt:1999kx,Antczak:2007fj} 
and theoretical%
\cite{Papanicolaou:1998uq,Montalenti:1999fj,Bogicevic:1999lr,Lorensen:1999qy}
work has been done.  Strain is known to govern the mesoscale morphology in
self-assembling systems\cite{Medhekar2007}.  While predictions about the
role of size and misfit for small islands go back over a
decade\cite{Hamilton:1996lr}, only recent experiments have captured and
quantified the rapid diffusion at ``magic sizes'' in the heteroepitaxial
Cu/Ag(111) system\cite{Signor2009}.  However, the experimental observations
of rapid diffusion from several distinct sizes of islands does not easily
comport with the simple model of Hamilton for magic sizes.  A missing
element in explaining the atomistic diffusion mechanism is the role of
island shape in controlling diffusion.  Understanding the trends of
diffusion barriers for small islands with island size and shape for the
common cubic (111) surface provides insight for experimental measurements
and thin-film growth, and provides the fundamental understanding to control
morphology.

Hamilton predicted a magic-size effect for heteroepitaxial islands with a
one-dimensional Frenkel-Kontorova model\cite{Aubry:1983lr} and a
corresponding two-dimensional equivalent\cite{Hamilton:1996lr}.  The 1D
model describes a chain of atoms harmonically coupled to their neighbors
and interacting with a rigid periodic substrate potential.  The lattice
misfit is the percentage difference between the equilibrium spring length
and the substrate periodicity, and the island misfit strain grows linearly
with the number of atoms in the island chain.  The diffusion barrier has
non-monotonic behavior with size, showing a minimum at the ``magic size''
equal to inverse of the lattice misfit.  At this size, the island
ground-state configuration contains one dislocation, where island atoms sit
at a peak of the substrate potential instead of a valley.  Islands below
the magic size are dislocation-free with a large energy barrier to nucleate
and propagate a dislocation for island diffusion, while larger islands
contain a dislocation and require an increasing energy barrier to move this
dislocation for island diffusion.  Using the embedded atom method
(EAM\cite{Foiles:1986fj}) for Ag/Ru(0001), Hamilton considered a 2D
equivalent with closed-shell islands, and the magic size corresponds with
the ground-state configuration containing one
dislocation\cite{Hamilton:1996lr}.  The ground state has atoms displaced
from FCC to HCP sites on the hexagonal Ru(0001) surface.  The dislocation
line marks the separation between FCC and HCP sections of the island.
Diffusion for these islands proceeds as all atoms in the island
collectively glide to continuously propagate the dislocation.

Reptation---first proposed for small island diffusion in homoepitaxial
systems---relies on dislocation movement\cite{Chirita:1999fk}.  Unlike
Hamilton's collective glide mechanism, reptation proceeds as the motion of
an island section from FCC to HCP sites on the $(111)$ surface, forming a
dislocation where the island atomic bonds are stretched.  Diffusion is
completed after the remaining island section subsequently follows in the
same direction to complete the transition.  Therefore the reptation
dislocation propagates through sequential motion of island sections while
the collective glide dislocation propagates by the continuous motion of the
entire island.

We find that the reptation diffusion mechanism exhibits a magic-size effect
in Cu/Ag(111) controlled by island shape that explains experimental
observations of anomalous diffusion.  Using an optimized EAM potential with
molecular dynamics and the dimer method, we calculate island diffusion
barriers up to 14-atom islands.  The diffusion barriers of different island
sizes and shapes is a non-monotonic function of the island misfit strain,
and separates into simple groups depending on the geometry of the island.
The shape effect is explained by the continuity of bonds during diffusion.
We find that the reptation mechanism is competitive over the glide
mechanism for all islands except those with closed-shell shapes.  After
considering the modulating effect of island geometry and the competition
between the diffusion mechanisms, we find multiple magic sizes each
diffusing with a single dislocation.  By considering island shape, we
predict a series of rapidly diffusing islands, each as their own magic
size.

Our study of island diffusion relies on an EAM potential optimized for the
prediction of island geometries, energetics, and kinetics for
Cu/Ag(111)\cite{Wu:2009lr}.  The potential was optimized using monomer and
dimer DFT energies and geometries.  The optimized EAM predicts the DFT
monomer diffusion barrier and the DFT energy difference between all-FCC
trimers and all-HCP trimers.
%Both of which are not part of the fitting database and demostrate the 
%reliability of the potential for larger island.  
However the potential (and DFT) overestimates the monomer and dimer
diffusion barriers at 93 and 88meV compared to
experimental\cite{Morgenstern:2004uq} values of 65$\pm$9 and 73meV,
respectively.  This is a carryover from DFT which has been shown to
overestimate surface adsorption energies\cite{Stampfl:2005lr}.  Due to this
discrepancy, we present our calculated island diffusion barriers relative
to the monomer diffusion barrier $\text{E}^\text{monomer}_\text{diff}$ of
93meV.

We anneal to determine island ground-state structures, and use molecular
dynamics\cite{Plimpton:1995lr}, dimer-search\cite{Henkelman:1999lr}, and
nudged elastic band\cite{Mills:1994yq} method to find island diffusion
transitions and determine diffusion barriers.  Multiple annealing runs for
each island size reveal compact islands with all FCC-site ground-states for
small sizes while mixed FCC/HCP ground-states exist only for the 13- and 14-atom
islands, the largest in our study.  We run direct molecular dynamics at
high temperature (600K) over several nanoseconds to explore possible
transitions.  We also use the dimer method which randomly searches through
phase-space for possible transitions from a starting state.  Nudged elastic
band finds the minimum energy pathway between the starting and ending states
of discovered transitions and extracts the diffusion barrier.

Different island shapes are possible for each island size in 2D, while two
different sites on the (111) surface allows different energies for the same
shape.  The islands form compact ground-state configurations that maximize
atomic coordination and minimize island strain.  The triangular (111)
surface lattice contains two hexagonal sublattices---FCC and HCP---each
with lattice spacing equal to that of the Ag nearest neighbor distance
$\text{nn}_{\text{Ag}} = 2.89$\AA.  A FCC-site is surrounded by three
HCP-sites in the $\langle11\bar2\rangle$ directions, while the next nearest
FCC-sites are in the close-packed $\langle\bar110\rangle$ directions.  Cu
islands, with bulk nearest neighbor distance $\text{nn}_{\text{Cu}} =
2.56$\AA, experience large misfit strains if all the island atoms sat
exclusively on one sublattice.  However, since FCC-sites are closer to
neighboring HCP-sites, the total island strain can be reduced if some atoms
in the island sit in a mixed FCC-HCP (FH) bond with a lattice site distance
of only $\frac{1}{\sqrt{3}}\text{Ag}_{nn} = 1.67$\AA\ rather than the
longer FCC-FCC (FF) or HCP-HCP (HH) bonds.  For a system with an opposite
lattice mismatch we would expect the next nearest neighbor FH-bonds with a
lattice site distance of $\frac{2}{\sqrt{3}}\text{Ag}_{nn} = 3.34$\AA; this
is also the case in the Pt/Pt(111) system\cite{Chirita:1999fk}.

We measure the substrate strain in each island in terms of the island
misfits in \Fig{str_drop}.  We relax ground-state configurations of
different islands and calculate the surface strain as a function of
distance away from the island's center-of-mass.  The atomic strain tensor
for each surface Ag atom is an average over the change in all
nearest-neighbor vectors\cite{Gullett:2008lr}.  In \Fig{str_drop}a the
tensile normal strain---sum of $\epsilon_{xx}$ and $\epsilon_{yy}$---drops
off as the inverse squared radial distance from the island's center-of-mass
and scales with island size; we define the scaling coefficient as the
island misfit, which has units of area.  This island misfit with the
substrate varies with the size and shape, and in \Fig{str_drop}b, we find a
linear correlation with the island size.  We use the island misfit as a
measure of the strain in the island because it contains information about
both the island size and shape.  Other island growth studies used the a
related measure of the substrate stress instead of strain\cite{Pao:2006lr},
which is linearly related.  This measure of substrate strain accounts for
the effect of island shape, and helps relate the 2D system back to the
simpler 1D model.

\bfig
\includegraphics[width=\smallfigwidth]{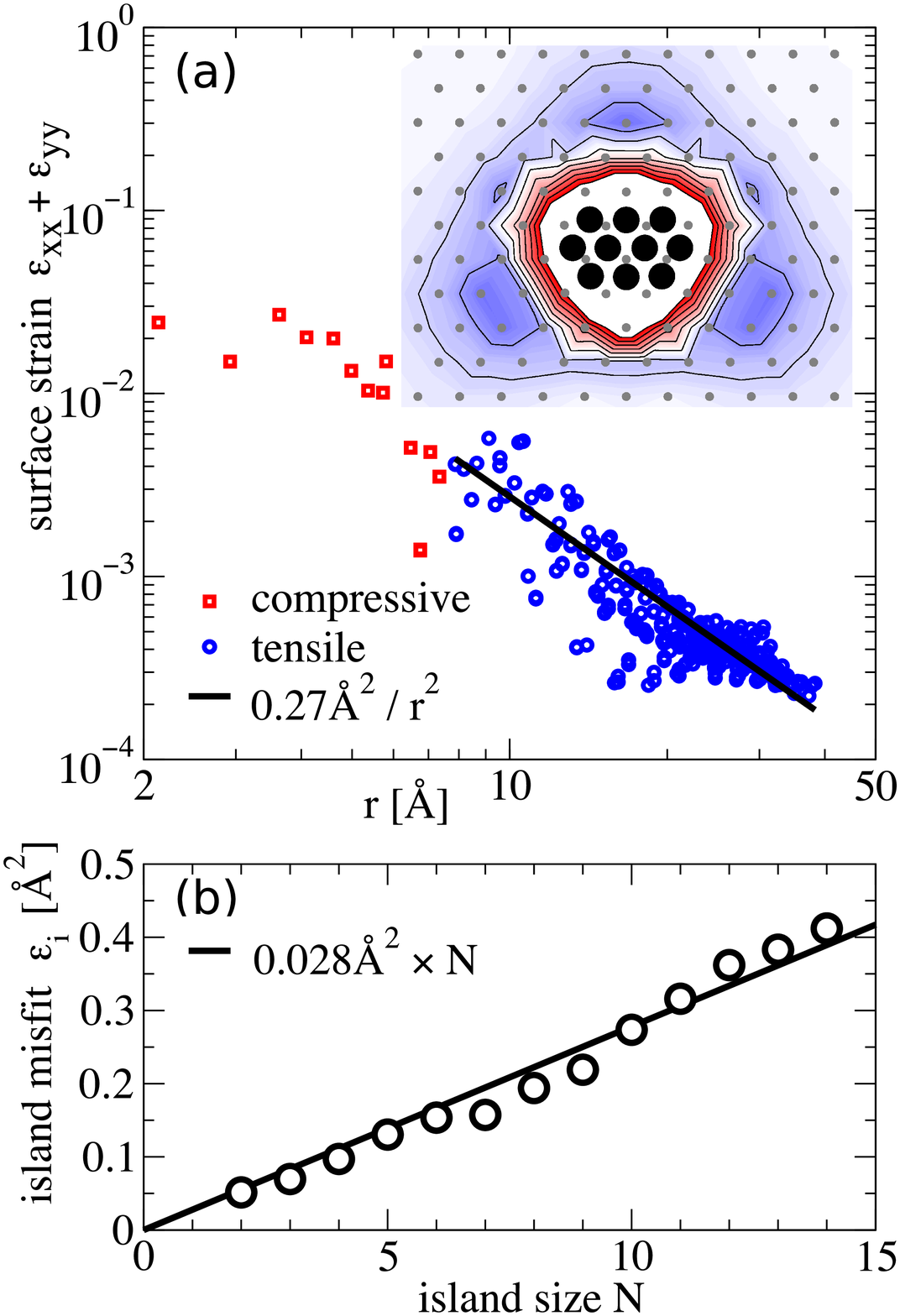}
\caption{Cu islands and strain in the Ag(111) surface.  
(a) Magnitude of the atomic strain plotted against the radial distance from
the 10-atom island center-of-mass.  The tensile strain
$\eps_{xx}+\eps_{yy}$ is the inverse squared radial distance times the
island misfit $\eps_\text{i}$.  The inset shows the hydrostatic strain
induced by the ground-state 10-atom Cu island in the Ag surface.
Compressive strain is in red (strains exceed --1.0\% under the island and
are not shown), and the compensating tensile strain is in blue.  (b) The
island misfit $\eps_\text{i}$ of ground-state island configurations grows
linearly with island size $N$ as 0.028\AA$^{2}$ $\times$ N, while
modulations show the influence of island shape.}
\label{fig:str_drop}
\efig
%%%%%%%%%%%%%%%%%%%%

Islands diffuse by the passage of a dislocation, with two different forms
in 2D: collective glide (\Fig{shear}b), or reptation (\Fig{shear}a).
Collective glide is the mechanism described by Hamilton, where the
dislocation propagates with the continuous motion of the entire island
moving from FCC to HCP sites, or vice versa.  Closed-shell configurations
(3-, 7-, and 12-atom islands) favor the collective glide mechanism, which
maintains neighboring bonds during diffusion.  Reptation involves
sequential motion of island sections to HCP sites, via a metastable
dislocated structure.  The FCC portion of the metastable state is separated
from the HCP portion by a dislocation with a $\langle\bar110\rangle$-type
line direction.  We identify this dislocation line direction to be characteristic 
of the reptation mechanism.  Island shape is
the critical criteria to determine whether a 2D island prefers the
collective glide mechanism or the reptation mechanism.

%%%%%%%%%%%%%%%%%%%%
\bfig
\includegraphics[width=\smallfigwidth]{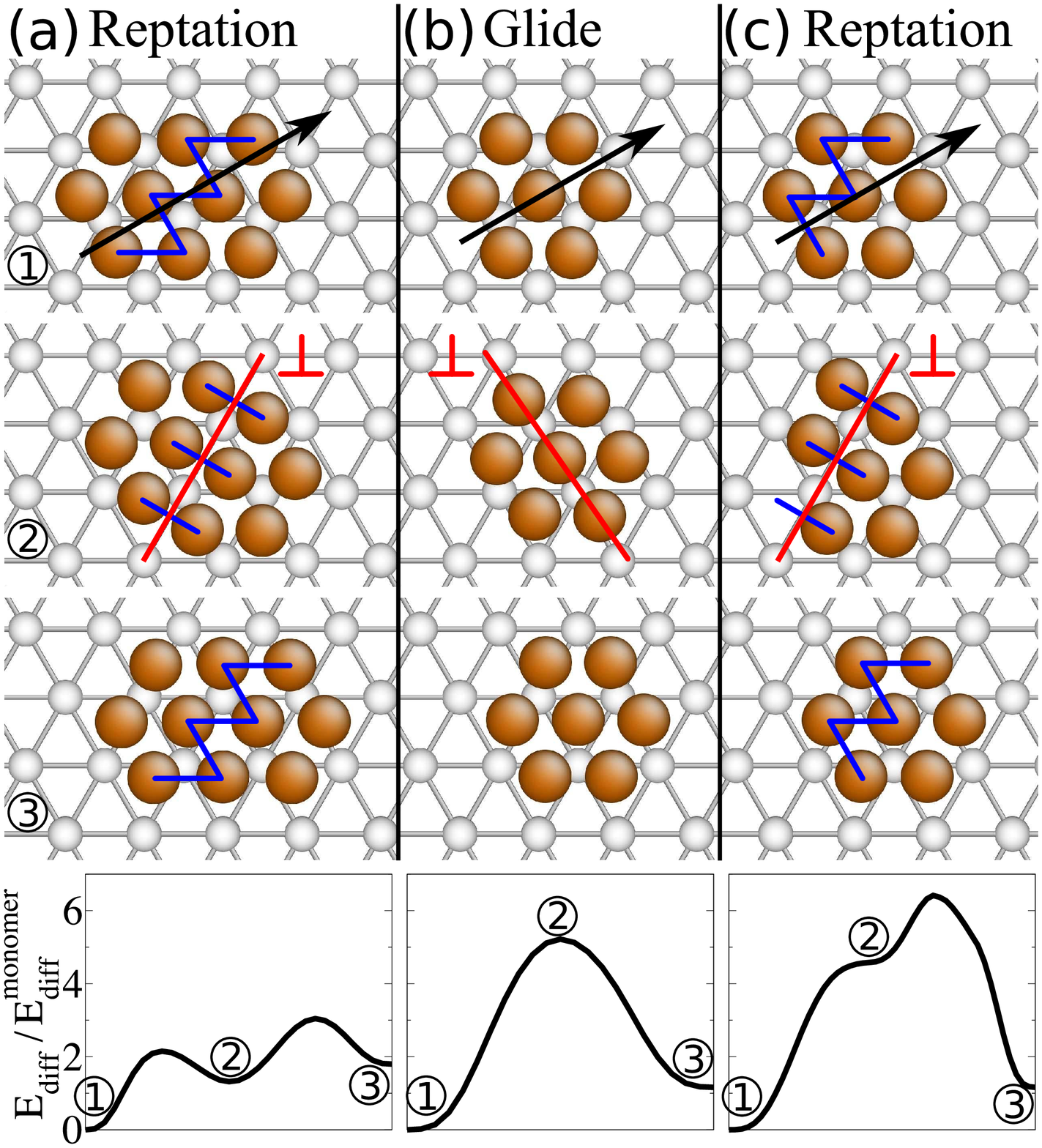}
\caption{Geometric requirements for low barrier island reptation diffusion.  
A $\bigtriangleup$ of Ag surround the FCC surface site and
$\bigtriangledown$ for the HCP surface site.  (a) An \textbf{allowed}
reptation transition involves transforming homogeneous bonds (1) into
heterogeneous bonds (2) \textit{without} leaving an unbonded Cu atom.  This
is a necessary condition for the reptation dislocation mediated island
diffusion.  (b) The collective glide dislocation mediated diffusion
mechanism observed for the 7-atom island.  Atoms shuffling in the direction
of diffusion propagates the passing dislocation (2).  (c) A
\textbf{disallowed} reptation transition for the 7-atom island leaves one
Cu atom without a bond across the dislocation line (2) and is high in
energy.  The energy barrier pathway for all three transitions is shown
below normalized with respect to the EAM monomer diffusion barrier of
93meV.}
\label{fig:shear}
\efig
%%%%%%%%%%%%%%%%%%%%

\Fig{shear} shows the geometric requirement to undergo the reptation
mechanism at a low energy barrier: no atom should be left without a bond
across the dislocation line.  The reptation mechanism slips part of the
island onto HCP sites where the number of $\langle\bar110\rangle$ island
rows sheared by the dislocation line form the same number of heterogeneous
FH-bonds (\Fig{shear}a.2).  The energy cost to form FH-bonds and move atoms
to HCP sites is compensated by the reduction of island strain from the
smaller island area of the dislocated state.  For the 10-atom island shown, the 
island misfit is reduced from 0.27\AA$^{2}$ (\Fig{shear}a.1) to 0.22\AA$^{2}$ 
(\Fig{shear}a.2).  An island with unequal number
of atoms across the dislocation line---two facing three in the 7-atom
island---has to overcome the additional barrier to break a bond during island shear
(\Fig{shear}c.2).  The dimer dissociation energy for Cu on Ag(111) is
$370\text{meV}\approx 4\text{E}^\text{monomer}_\text{diff}$, which
increases the reptation mechanism energy barrier for closed-shell islands
above the collective glide mechanism barrier.  Ultimately, the absence of
bond-breaking during slip is the deciding factor in allowing a low energy
reptation mechanism.

\Fig{str_v_barr} groups barriers for islands with equal numbers of
sheared $\langle\bar110\rangle$ rows, demonstrating the shape-modulated
magic-size effect.  For reptation-allowed islands, this groups islands with
the same number of FH-bonds in the dislocated state, and for
reptation-disallowed islands this is the number of $\langle\bar110\rangle$
rows in the diffusion direction.  The 2- and 3-row groups appear very
similar to the characteristic magic-size effect plot for 1D island chains
diffusion\cite{Hamilton:1996lr}.  The 7- and 8-atom island configurations
in the 2-row group are not ground-states and are constructed to test the
continuation of the 2-row magic-size effect.  The shape of these two
islands induce larger strains than in their ground-states and promotes the
dramatic reduction in diffusion barrier.  The shape effect from sheared
$\langle\bar110\rangle$ rows gives different magic-size regions even though
only one dislocation is present in all cases.

%%%%%%%%%%%%%%%%%%%%
\bfig
\includegraphics[width=\smallfigwidth]{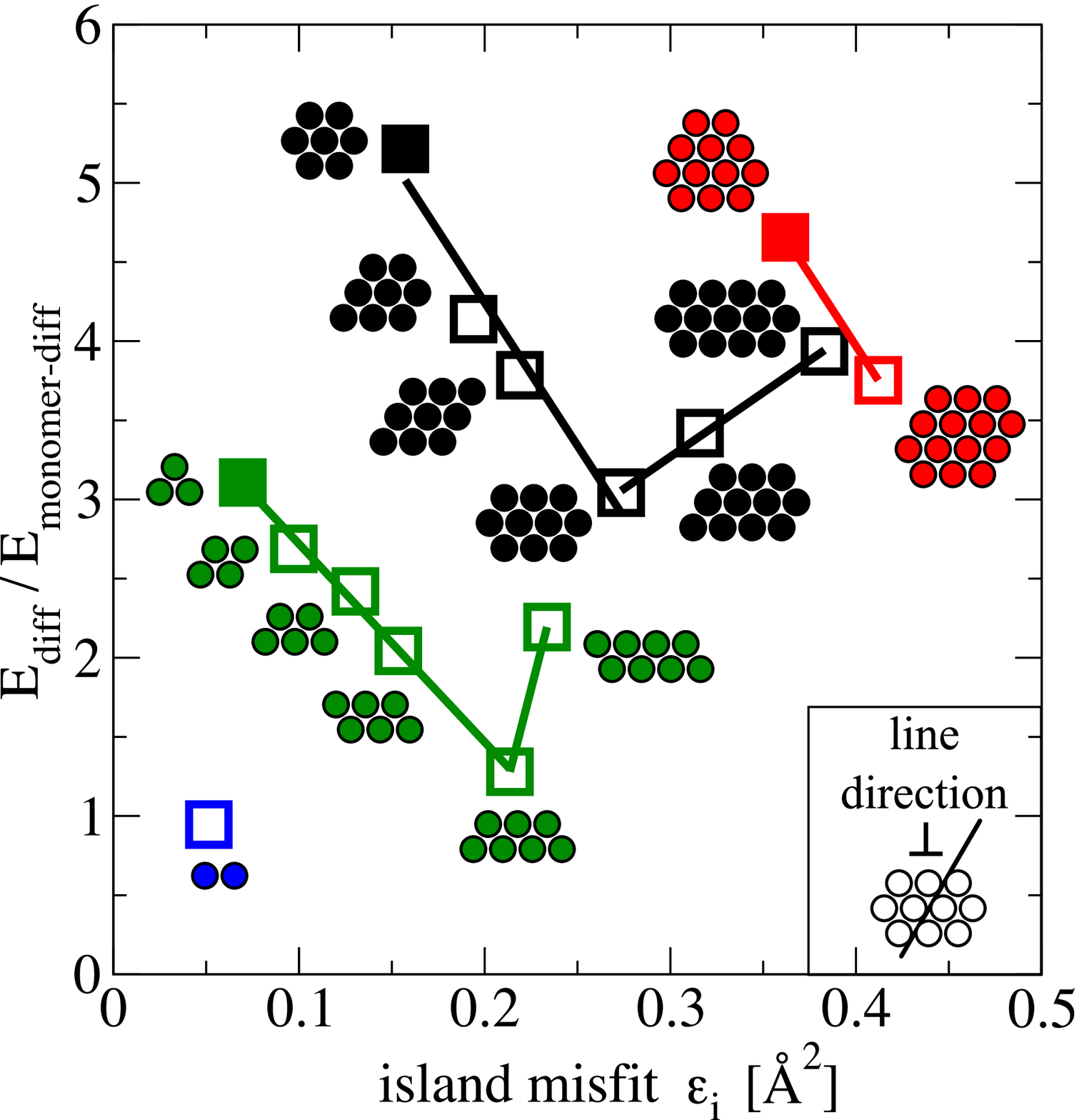}
\caption{The island diffusion barriers versus island misfit $\eps_\text{i}$
for different islands showing a shape-modulated magic-size effect.  The
diffusion barrier is normalized with respect to the EAM monomer diffusion
barrier of 93meV.  The open symbols represent islands that allow reptation
according to their shape, while the filled symbols are islands that do not
(the configurations shown are oriented with respect to the dislocation line
direction in the inset).  Islands with the same number of sheared
$\langle\bar110\rangle$ rows along the dislocation line follow a ``magic
size'' trend in barriers as a dislocation aids in diffusion---with blue for
1-, green for 2-, black for 3-, and red for 4-rows.  The distinct groupings
of magic-size regions highlight the key role of shape in the diffusion of
small islands with a large lattice misfit.  The 7- and 8-atom 2-row
configurations in green are not energetically favorable, but show the
continuation of the 2-row magic-size effect.  Note: The glyphs shown for
the 6-, 9- and 11-atom islands are not the ground-state configurations for
that size, but instead represent the shape before a diffusion transition.}
\label{fig:str_v_barr}
\efig
%%%%%%%%%%%%%%%%%%%%

The transition pathways for 2D islands of different row-groups in
\Fig{str_v_barr} also follow the trends seen for the 1D island chains.  In
the 2-row group, the dislocated state is not metastable for the 4-atom
island and becomes more and more stable with increasing island size.  The
dislocated state of the 7-atom 2-row is almost equal in energy with the
undislocated state, while the dislocated state of the 8-atom 2-row is the
ground state and possesses a higher diffusion barrier.  In the 3-row group,
the dislocated state is not metastable for the both the 8- and 9-atom
islands and becomes metastable at 10-atoms.  Continuing the 3-row group,
the dislocated state of the 11-atom island is still metastable, but the
diffusion barrier is higher due to asymmetric island structure.  Finally,
the dislocated state for the 13-atom island is the ground-state and the
diffusion barrier becomes even higher.  The minima for ground-state
diffusion barriers---6- and 10-atom islands---corresponds well with
experimental observations, as well as the immobility of 7-atom
islands\cite{Signor2009}.

Island shape controls the 2D magic-size effect for Cu/Ag(111) where a
combination of island geometry and misfit produces multiple magic sizes
even for single dislocation-mediated diffusion.  We find that the
reptation diffusion mechanism allows for greatly reduced diffusion barriers
for heteroepitaxial systems compared with the collective glide mechanism.
The criteria for the reptation mechanism requires the island shape to be
such that no atom is left unbonded across the dislocation line.  This
mechanism predicts multiple magic sizes even for small ($<20$) island
sizes, which quantitatively agrees with experimental observations of
Cu/Ag(111).  We expect similar effects in other heteroepitaxial systems
with large lattice misfits, and for the magic size islands to affect the
growth morphologies for low coverages.

\begin{acknowledgments}
The authors thank John Weaver for helpful discussions.  This research is
supported by NSF/DMR grant 0703995, and 3M's Untenured Faculty Research
Award.
\end{acknowledgments}

%%%%%%%%%%%%%%%%%%%%%%%%%%%%%%%%%%%%%%%%%%%%%%%%%%%%%%%%%%%%%%%%%%%%%%%%

\end{document}